\documentclass[10pt]{article}
\usepackage{amssymb}
\usepackage{amsmath}
\usepackage{wasysym}
\usepackage{graphicx}
\begin{document}
	
\title{Non-separable wave evolution equations in quantum kinetics}

 \author{C. Dedes\thanks{{c\_dedes@yahoo.com}}\\
Workwell Solutions\\
Radius House,  51 Clarendon Road, \\
Watford, Hertfordshire, \\
WD17 1HP,\\
United Kingdom \\ }

  \maketitle

\begin{abstract}

A non-separable wave-like integro-differential equation for the time evolution of the Wigner distribution function in phase space is educed from the corresponding separable kinetic equation. It is shown that it leads to non-local dispersion effects that may involve complex group velocities and also display non-unitary features as entropy production. By employing the quantum hydrodynamical description a non-local evolution wave equation is also derived by synthesizing the Hamilton-Jacobi equation with that of continuity, which predicts the generation of quadrupole quantum effects in the propagation of the spatial probability density. Extension of the formalism for a boson scalar field is also presented along with a brief commentary on the issue of irreversibility.

\end{abstract}

\section{Introduction}
 
The philosophical underpinnings of quantum theory are related as a rule to the wavefunction, a theoretical device with a rather controversial ontological status, abound with intricate metaphysical issues. Yet one might argue that the great conceptual novelty of quantum theory, namely its irreducibly non-separable character, is introduced in the framework of the theory somehow externally, through the use of the \textit{ad hoc} ansatz of the entangled states. It must be noted that in the Bohmian formulation, the appearance of non-separability and non-locality is also expressed with the help of the notion of quantum potential \cite{Bohm, Cushing}. Another issue that needs to be addressed is the incorporation of second-order time derivatives in the theory. Historically, a partial differential equation of this general type for the time evolution of the wave function had been put forward by Schr\"{o}dinger to avoid the possibility of complex-valued wavefunctions \cite{Callender}. Schr\"{o}dinger also admitted the possibility of establishing governing equations for the probability density which is directly observable instead of the wave function yet he did not consider it at greater length as it would lead to immense complications (\cite{Schrodinger}- see also the related discussion in \cite{Dyson}). Here instead we seek a formalism that attempts to incorporate the notion of non-separability in a more pronounced way and to achieve this we employ the phase space description of quantum mechanics and deduce a second order in time governing equation from the familiar transport equation of the Wigner function. We also repeat the same procedure for other transport equations used in physical kinetics and non-equilibrium systems and examine some of the consequences. It is found that non-local dispersion effects may be exhibited both in space and time and for certain quantum distribution functions, as the propagated probability density in the configuration space, quadrupole phenomena may arise.

\section{Wave equations in phase space}

It is well known that an autonomous formulation of quantum mechanics is possible through the use of quasi-probability density functions in phase space \cite{Wigner, Case}. The phase-space description is based on Wigner's probability distribution which is real but it can be negative, which is generally considered as a sign of non-classicality. From its marginal integrals, we can calculate the positive definite probability densities in the configuration or the momentum space. The free two-particle Wigner function \cite{Lee, Bell} in particular $ W_{12}(x_{1},x_{2},p_{1},p_{2},t)$ evolves according to the equation of motion

\begin{equation}
   \left (\frac{\partial}{\partial t}+\frac{p_{1}}{m_{1}}\frac{\partial }{\partial x_{1}}+\frac{p_{2}}{m_{2}}\frac{\partial }{\partial x_{2}}\right) W_{12}=0.
\end{equation}

\noindent
By making an appropriate transformation the above equation of motion indicates a spreading of the Wigner function which remains positive \cite{Lee}. This resembles a classical equation, yet one may argue that even without interactions, non-classical correlations may be present. In our view, it would be a \textit{desideratum} of the theoretical framework to arrive at a manifestly non-classical free dynamical equation for the two-particle case. To do this we take the time derivative and consequently the corresponding two spatial derivatives and then subtract, or equivalently multiply (1) with 

\begin{equation}
\left (\frac{\partial}{\partial t}-\frac{p_{1}}{m_{1}}\frac{\partial }{\partial x_{1}}-\frac{p_{2}}{m_{2}}\frac{\partial }{\partial x_{2}}\right) W_{12}=0 ,\\
\end{equation}

\noindent
and we arrive at the wave evolution equation for the free two-particle Wigner function $W_{12}=W_{12}(x_{1},x_{2},p_{1},p_{2},t)$  

\begin{equation}
   \left(\frac{\partial ^{2}}{\partial t^{2}}-
     \frac{p^{2}_{1}}{m^{2}_{1}}\frac{\partial ^{2}}{\partial x_{1}^{2}}- \frac{p^{2}_{2}}{m^{2}_{2}}\frac{\partial ^{2}}{\partial x_{2}^{2}}- \frac{2p_{1}p_{2}}{m_{1}m_{2}}\frac{\partial ^{2}}{\partial x_{1}\partial x_{2}}\right)W_{12}=0.
\end{equation}

\noindent
This suggests that even without any potentials present, non-local correlations may be induced. The presence of both position and momentum may seem more appropriate for settings similar to Einstein-Podolsky-Rosen \cite{EPR} or Popper's thought experiment \cite{Shih} as it indicates non-separability and even modulation of the momentum diffraction of one particle in space-like separated regions. The kind of inter-site spatial derivatives that appear in (3) can also be found in the non-relativistic Bethe-Salpeter equation \cite{Schweber}, the Pauli-Jordan commutator relations \cite{Bohr} and the Wigner evolution equation that includes potential interactions \cite{Scully}; in contradistinction, the two particles are free in our formula. To grasp more adequately the physical meaning of the dynamic evolution equation we employ the field description and notice that we can obtain the equation of motion (3) through the Lagrangian density

\begin{equation}
    L=\frac{1}{2}\left[\left(\frac{\partial W}{\partial t}\right)^{2}-\frac{p^{2}_{1}}{m^{2}_{1}}\left(\frac{\partial W}{\partial x_{1}}\right)^{2}-\frac{p^{2}_{2}}{m^{2}_{2}}\left(\frac{\partial W}{\partial x_{2}}\right)^{2}-\frac{2p_{1}p_{2}}{m_{1}m_{2}}\frac{\partial W}{\partial x_{1}}\frac{\partial W}{\partial x_{2}}\right],
\end{equation}

\noindent
which indicates an explicitly non-local wave field density through the last potential term. 

Transforming (3) to the centre of mass and relative distance coordinates

\begin{gather}
    X=\frac{m_{1}x_{1}+m_{2}x_{2}}{m_{1}+m_{2}} \\
  \chi=x_{1}-x_{2},
\end{gather}

\noindent
we obtain

\begin{align}
   & \Bigg\{\frac{\partial ^{2} }{\partial t^{2}}-\left(\frac{p_{1}+p_{2}}{m_{1}+m_{2}}\right)^{2}\frac{\partial ^{2}}{\partial X^{2}}-\left(\frac{p_{1}}{m_{1}}-\frac{p_{2}}{m_{2}}\right)^{2}\frac{\partial ^{2}}{\partial \chi^{2}} \nonumber\\
   & +\frac{2}{m_{1}+m_{2}}\left[\frac{p_{1}^{2}}{m_{1}}-\frac{p_{2}^{2}}{m_{2}}-\frac{p_{1}p_{2}}{m_{1}m_{2}}\left(m_{1}-m_{2}\right)\right]\frac{\partial ^{2}}{\partial \chi \partial X}\Bigg\}W_{12}=0.
\end{align}

\noindent
By choosing the sum of the two-particle momenta to be correlated in an EPR fashion

\begin{gather}
   p_{1}+p_{2}=0  , \nonumber\\
   p_{1}=-p_{2}=p. \nonumber\\
\end{gather}

\noindent
This can be reduced to 

\begin{equation}
 \left[\frac{\partial ^{2} }{\partial t^{2}}-\left(\frac{p}{\mu}\right)^{2}\frac{\partial ^{2}}{\partial \chi^{2}} \right]W_{12}=0, 
\end{equation}

\noindent
where $\mu$ is the reduced bi-particle mass. The corresponding dispersion relation $\omega=\frac{p}{\mu}k$ and the wave propagation velocity defined by is $\frac{p}{\mu}$. This is a standard one-dimensional wave equation that admits solutions of the form

\begin{equation}
    W_{12}(x_{1}-x_{2},t)=f\left(x_{1}-x_{2}-\frac{p}{\mu}t\right)+g\left(x_{1}-x_{2}+\frac{p}{\mu}t\right),
\end{equation}

\noindent
assuming that the initial conditions for the first order time derivative is zero. Another solution has a cosine squared dependence and this would effectively generate the collapse and revival of the probability density. In contrast, the free-particle equation (1) after a transformation of the centre of mass and relative distance coordinates admits solutions similar to the first term only. 

The Wigner function $W(\mathbf{r},\mathbf{p},t)$ for a single particle problem in the presence of a potential $V(x)$ is governed by the equation

\begin{equation}
    \frac{\partial W}{\partial t}+\frac{\mathbf{p}}{m}\cdot\nabla W=\frac{2}{\hbar}\sin\left(\frac{\hbar}{2}\nabla_{\mathbf{r}}\nabla_{\mathbf{p}}\right)VW,
\end{equation}

\noindent
where the spatial derivative $\nabla_{\mathbf{r}}$ acts only on the potential energy function and the momentum gradient $\nabla_{\mathbf{p}}$ operates on the Wigner function only. Multiplying (11) with 

\begin{equation}
    \frac{\partial }{\partial t}-\frac{\mathbf{p}}{m}\cdot\nabla_{\mathbf{r}}, \nonumber\\
\end{equation}

\noindent
and assuming the potential is time-independent we find

\begin{align}
    \Box^{2} W=& \frac{2m}{\hbar c^{2}}\sin\left(\frac{\hbar}{2}\nabla_{\mathbf{r}}\nabla_{\mathbf{p}}\right)\left[\left(\frac{\mathbf{p}}{m}\cdot\nabla_{\mathbf{r}}V\right)W+\left(\frac{2\mathbf{p}}{m}\cdot\nabla_{\mathbf{r}}W\right)V\right] \nonumber\\
     & +\frac{4}{\hbar ^{2}c^{2}}\sin \left(\frac{\hbar}{2}\nabla_{\mathbf{r}}\nabla_{\mathbf{p}}\right)V\sin \left(\frac{\hbar}{2}\nabla_{\mathbf{r}}\nabla_{\mathbf{p}}\right)VW,
\end{align}

\noindent 
where the velocity of propagation in the d'Alembertian operator $\Box^{2}$ is defined as earlier so $c^{2}=\frac{p^{2}}{m^{2}}$.

By including time-dependent interaction terms, the $N$ particle three-dimensional generalization of (1) is presented in an integro-differential form similar to the Boltzmann equation as

\begin{equation}
      \left[\frac{\partial }{\partial t}+\left(\sum_{i=1}^{N}\frac{\mathbf{p}_{i}}{m_{i}}\cdot\nabla_{i}\right)\right]W=\int d^{N}\mathbf{p}'\tilde{V} W,
\end{equation}

\noindent
where $\tilde{V}=\tilde{V}(\mathbf{r}_{1},..\mathbf{r}_{N};\mathbf{p}_{1}-\mathbf{p}'_{1},.,\mathbf{p}_{N}-\mathbf{p}'_{N})$ is the the Wigner potential and $W=W_{1..N}$ the multi-particle quasi-distribution function. We also define the total time derivative,

\begin{equation}
   \frac{d}{dt} =\frac{\partial}{\partial t}+\sum_{i=1}^{N}\frac{\mathbf{p}_{i}}{m_{i}}\cdot\nabla_{i}.
\end{equation}

\noindent
In this notation, the transport equation for the Wigner function can be written simply as 

\begin{equation}
  \dot{W}=\int  d^{N}\mathbf{p}'\tilde{V} W , 
\end{equation}

\noindent
and repeating the same procedure as we did earlier which effectively multiplies both sides of (13) with 

\begin{equation}
\frac{\partial}{\partial t}-\sum_{i=1}^{N}\frac{\mathbf{p}_{i}}{m_{i}}\cdot\nabla_{i} ,\nonumber\\
\end{equation}

\noindent
yields

\begin{equation}
    \left(\frac{\partial}{\partial t}-\sum_{i=1}^{N}\frac{\mathbf{p}_{i}}{m_{i}}\cdot\nabla_{i} \right)\left(\frac{\partial}{\partial t}+\sum_{i=1}^{N}\frac{\mathbf{p}_{i}}{m_{i}}\cdot\nabla_{i} \right)W
   = \int d^{N}\mathbf{p}' \hat{U}W,
\end{equation}

\noindent
where

\begin{equation}
    \hat{U}=\frac{\partial \tilde{V}}{\partial t}-\sum_{i=1}^{N}\frac{\mathbf{p}_{i}}{m_{i}}\cdot\nabla_{i}\tilde{V} 
    -\tilde{V}\left(\sum_{i=1}^{N}\frac{2\mathbf{p}_{i}}{m_{i}}\cdot\nabla_{i}+\int d^{N}\mathbf{p}' \tilde{V} \right).
\end{equation}

\noindent
The above time-symmetric integro-differential equation (16) is the central result presented in this work. On the left-hand side, it includes $N(N-1)/2$ terms with spatial derivatives acting simultaneously at distant sites which reflects the irreducible nature of quantum non-separability and under appropriate conditions may dominate over the separable terms. In the presence of vector potentials, a product of covariant derivatives acting simultaneously at two distant sites appears in place of the spatial derivatives and results also in products of vector potentials in the equation of motion. 

The condition that must be fulfilled so that (16) will also verify (13) is

\begin{equation}
  \left(\int  d^{N}\mathbf{p}'\tilde{V} W  \right)^{2}=  2\int dW \int  d^{N}\mathbf{p}'\left( \dot{\tilde{V}}W +\tilde{V}\int d^{N}\mathbf{p}' \tilde{V}W \right),
\end{equation}

\noindent
but this is not always the case as we will see. Since this is a second-order in time differential equation it must be supplemented by initial conditions for the first-order time derivative, provided from (13), and this entails the possibility of discrepancy between (16) and (13) which means that a solution to the latter will not correspond to a wavefunction and cannot be guaranteed that the probability density associated with it will be positive definite. It must be added at this point that the existence of negative probabilities does not by itself invalidate a physical theory but rather points to a fundamental unattainability in establishing the necessary assumed conditions for its verification \cite{Feynman}. Interestingly, this argumentation is reminiscent of Bohr's reply to EPR. The physical meaning then of this kind of solutions is itself an interesting issue but the main claim of this paper is that even standard solutions need to satisfy a non-separable equation. For the single-particle case (16) reduces to the following integro-differential wave equation

\begin{equation}
\Box ^{2}W =\frac{1}{c^{2}}\int d\mathbf{p}' \left[\tilde{V}\left(\frac{2\mathbf{p}}{m}\cdot\nabla+\int d\mathbf{p}' \tilde{V} \right)-\left(\frac{\partial \tilde{V}}{\partial t}-\frac{\mathbf{p}}{m}\cdot\nabla\tilde{V}\right)\right]W.
\end{equation}

\noindent
which is an alternate form of (11).

By employing the total time derivative of the equation of motion for the Wigner distribution we can write down a convective wave equation like

\begin{equation}
    \ddot{W}-\sum_{i=1}^{N}\frac{2\mathbf{p}_{i}}{m_{i}}\cdot\nabla_{i}\dot{W}=\int d^{N}\mathbf{p}'\Bigg[\left( \dot{\tilde{V}}-\sum_{i=1}^{N}\frac{2\mathbf{p}_{i}}{m_{i}}\cdot\nabla_{i}\tilde{V}\right) -\tilde{V}\left(\sum_{i=1}^{N}\frac{\mathbf{p}_{i}}{m_{i}}\cdot\nabla_{i}+\int d^{N}\mathbf{p}' \tilde{V} \right)\Bigg] W,
\end{equation}

\noindent
or, formally another variant of (20) is

\begin{align}
   & \dot{W}^{2}= 2\int dW \Bigg\{\sum_{i=1}^{N}\frac{2\mathbf{p}_{i}}{m_{i}}\nabla_{i}\dot{W}+\int d^{N}\mathbf{p}'\Bigg[\left( \dot{\tilde{V}}-\sum_{i=1}^{N}\frac{2\mathbf{p}_{i}}{m_{i}}\cdot\nabla_{i}\tilde{V}\right)\nonumber\\
& -\tilde{V}\left(\sum_{i=1}^{N}\frac{\mathbf{p}_{i}}{m_{i}}\cdot\nabla_{i}+\int d^{N}\mathbf{p}' \tilde{V} \right)\Bigg]W+C\Bigg\},
\end{align}

\noindent
where $C$ is an integration constant. This is non-linear so obviously, it violates the superposition principle. We can then identify two distinct solutions, one forward evolving one and a backward. It must be noted that each of these solutions is time-asymmetric, since the collision integral that follows from the square root of (21) is not invariant under time reversal, which in turn introduces to the formalism an element of genuine irreversibility. 
Approximately it holds that

\begin{equation}
    \dot{W}^{2}\approx  2\int dW \int  d^{N}\mathbf{p}'\left( \dot{\tilde{V}}W +\tilde{V}\int d^{N}\mathbf{p}' \tilde{V}W \right).
\end{equation}

 It is also evident that (16) is invariant under particle exchange from which it follows that the probability density is also invariant.
\begin{equation}
    \hat {P}W_{12..m..n..N}=W_{12..n..m..N} \Rightarrow \hat {P}\rho _{12..m..n..N}=\rho _{12..n..m..N}.
\end{equation}

We may extend the theory to describe scalar fields, considering a free bosonic field $\Phi$, described by an Schr\"{o}dinger functional $\Psi[\Phi]$ \cite{Hatfield}; accordingly, the transport equation for the Wigner functional of this field is \cite{functional, functional2}

\begin{equation}
\left(\frac{\partial }{\partial t}+\int dx \dot{\Phi}\frac{\delta}{\delta \Phi}\right)W[\Phi, \dot{\Phi};t ]=\int dx \nabla ^{2}\Phi \frac{\delta W[\Phi, \dot{\Phi};t ]}{\delta \dot{\Phi}}.
\end{equation}

\noindent
Multiplying both sides of the above with 

\begin{equation}
    \frac{\partial }{\partial t}-\int dx \dot{\Phi}\frac{\delta}{\delta \Phi}, \nonumber\\
\end{equation}

\noindent
we obtain a functional wave-like differential equation of the form

\begin{align}
 & \Bigg\{\left[\frac{\partial ^{2}}{\partial t ^{2}}-\int dx \dot{\Phi}\int dx \left(\frac{\delta}{\delta \Phi}+\dot{\Phi}\frac{\delta ^{2}}{\delta \Phi ^{2}}-\nabla^{2}\Phi \frac{\delta ^{2}}{\delta \Phi \delta \dot{\Phi}}\right)\right]- \nonumber\\
 & \int dx \nabla ^{2} \Phi \int dx \left(\frac{\delta}{\delta \Phi}+\dot{\Phi}\frac{\delta ^{2} }{\delta \Phi \delta \dot{\Phi}}+\nabla ^{2}\Phi \frac{\delta ^{2}}{\delta \dot{\Phi}^{2}}\right)\Bigg\}W[\Phi, \dot{\Phi};t ]=0.
\end{align}

\section{Non-local probability density wave equation and quadrupole effects}

We may repeat a similar procedure starting from the continuity equation for the propagated probability density, which in tensor notation is written as

	\begin{equation}
	\frac{\partial \rho}{\partial t} +v_{i}\frac{\partial \rho }{\partial x_{i}} =-\rho \frac{\partial v_{i}}{\partial x_{i}}. 
	\end{equation}

 \noindent
 where $v$ the hydrodynamic velocity
	
	\begin{equation}
	v=\frac{1}{m}\frac{\partial S}{\partial x_{i}}.
	\end{equation}

\noindent
The Hamilton-Jacobi is expressed as

	\begin{equation}
	- \frac{\partial S}{\partial t}-\frac{1}{2m}\left(\frac{\partial S}{\partial x_{i}}\right)^{2}=V+Q.
	\end{equation}
	
\noindent
where 
	
	\begin{equation}
	Q=-\frac{\hbar^{2}}{4m}\left[\frac{1}{\rho}\frac{\partial ^{2}\rho}{\partial x_{i}^{2}}-\frac{1}{2\rho^{2}}\left(\frac{\partial \rho}{\partial x_{i}}\right)^{2}\right]  
	\end{equation}

 \noindent
 the quantum potential \cite{Bohm} which adds a correction term of quantum origin to the equation of motion,

	\begin{equation}
	\frac{du_{i}}{dt}=\frac{\partial v_{i}}{\partial t}+v_{j}\frac{\partial v_{i}}{\partial x_{j}}=-\frac{1}{m}\frac{\partial} {\partial x_{i}} (V+Q).
	\end{equation}

 \noindent
The Lagrangian derivative which expresses the total time derivative concerning time for a particle following the probability fluid motion is $\frac{d}{dt}=\frac{\partial }{\partial t}+v_{j}\frac{\partial }{\partial x_{j}}$ and $\frac{\partial }{\partial t}-v_{j}\frac{\partial }{\partial x_{j}}$ is the corresponding expression for a particle moving with the opposite velocity. The momentum equation is
	
	\begin{equation}
	\frac{\partial  (\rho v_{i})}{\partial t}=-\frac{1}{m}\partial _{x} \Pi-\frac{\rho}{m} \partial_{x} V,
	\end{equation}

 \noindent	
and the stress tensor is written as \cite{Harvey}

\begin{equation}
    \Pi_{ij}=\frac{\hbar ^{2}}{4m}\left(\delta _{ij}\nabla ^{2}\rho-\nabla _{i}\rho\nabla _{j}ln\rho\right).
\end{equation}	

\noindent
Multiplying (26) with $\frac{\partial }{\partial t}-v_{j}\frac{\partial }{\partial x_{j}}$ gives

	 \begin{equation}
    \frac{ \partial^{2} \rho}{\partial t^{2}}=-\frac{\partial}{\partial x_{i}}\cdot \frac{\partial (\rho v_{i})}{\partial t},
 \end{equation}

 \noindent
 or,

 \begin{equation}
\frac{\partial ^{2}\rho}{\partial t^{2}}=\frac{\partial ^{2}(\rho u_{i}u_{j})}{\partial x_{i}\partial x_{j}}-\rho \frac{du_{i}}{dt}.
\end{equation}

 \noindent
 More compactly this is written as

 \begin{equation}
     \frac{\partial ^{2}\rho}{\partial t^{2}}=\frac{\partial ^{2}T_{ij}}{\partial x_{i}\partial x_{j}},
 \end{equation}

\noindent
where $T_{ij}$ a quantum Lighthill tensor

	\begin{equation}
    T_{ij}=\rho u_{i}u_{j}-\Pi_{ij}-c_{0}^{2}\rho\delta _{ij},
\end{equation}

\noindent
that generates dipole and quadrupole effects and also includes a subtracted reference velocity term $c_{0}$. The same single-particle equation can be derived using the same methodology as in the derivation of the Lighthill equation in aeroacoustics \cite{Goldstein}. Yet, for the many-particle cases, the two prescriptions lead to different results. The formal solution of this inhomogeneous wave equation is

\begin{equation}
    \rho = \frac{1}{4\pi c_{0}^{2}}\frac{\partial ^{2} }{\partial x_{i}\partial x_{j} }\int \frac{ T_{ij}\left(t-\frac{|\mathbf{x}-\mathbf{x}'|}{c_{0}}\right)}{|\mathbf{x}-\mathbf{x}'|}d^{3}\mathbf{x}'-\frac{1}{4\pi c_{0}^{2}}\frac{\partial}{\partial x_{i}} \int _{S}\frac{1}{m}\rho \nabla V d^{3}\mathbf{x}',
\end{equation}

\noindent
and in the far-field zone,

\begin{equation}
    \rho \approx \frac{x_{i}x_{j}}{4\pi c_{0}^{4}x^{3}}\frac{\partial ^{2} }{\partial t^{2}}\int \frac{ T_{ij}\left(t-\frac{|\mathbf{x}-\mathbf{x}'|}{c_{0}}\right)}{|\mathbf{x}-\mathbf{x}'|}d^{3}\mathbf{x}'+\frac{x_{i}}{4\pi c_{0}^{3}x^{2}}\frac{\partial}{\partial t} \int _{S}\frac{1}{m}\rho \nabla V d^{3}\mathbf{x}'.
\end{equation}

We may account for spin effects by employing the Pauli-Schr\"{o}dinger Hamiltonian which acts on a spinor wavefunction where $\theta,\phi$ the Euler angles and $\mathbf{A}$ the vector potential, gives the following expression for the velocity \cite{Bohm, Cushing}

\begin{equation}
    v=\frac{1}{2m}\left(\nabla S+cos\theta\nabla\phi\right)-\frac{q}{m}\mathbf{A},
\end{equation}

\noindent
and the Hamilton-Jacobi and the wave equation for the probability density can be derived as previously.

\textit{Modification of the continuity equation for a bosonic field} The case of quantum fields is a bit more involved. We can expand the presented theoretical structure for a spin-zero scalar field by substituting the wave-functional into the Schwinger-Tomonaga equation. The Hamilton-Jacobi equation reads as

\begin{equation}
   \frac{\partial S}{\partial t}+Q+\frac{1}{2}\int d^{3}x\left[\left(\frac{\delta S}{\delta \phi}\right)^{2}+(\nabla\phi)^{2}\right] =0 ,
\end{equation}

\noindent
where

\begin{equation}
    Q=-\frac{1}{2}\int d^{3}x \frac{1}{R}\frac{\delta ^{2} R}{\delta \phi ^{2}},
\end{equation}

\noindent
the super quantum potential and $\rho=R^{2}$ the functional probability density. The evolution of this density is governed by the continuity equation

\begin{equation}
  \frac{\partial \rho}{\partial t} +\int d^{3}x \frac{\delta }{\delta \phi}\left(\rho\frac{\delta S}{\delta \phi}\right)=0 
\end{equation}

\noindent
Motivated by the many-particle cases and using \cite{Bohm}

\begin{equation}
    \Box ^{2}\phi=-\frac{\delta Q}{\delta \phi},
\end{equation}

\noindent
we obtain

\begin{equation}
    \ddot{\rho}-2\int d^{3}x \frac{\partial \phi}{\partial t}\frac{\delta \dot{\rho}}{\delta \phi} \approx \int d^{3}x \left(\rho \frac{\delta ^{2} Q}{\delta \phi ^{2}}-\dot{\rho}\frac{\delta ^{2} S}{\delta \phi ^{2}}+2\int d^{3}x\frac{\partial \phi}{\partial t}\frac{\delta \rho}{\delta \phi}\frac{\delta ^{2}S}{\delta \phi ^{2}}\right),
\end{equation}

\noindent
in which third-order functional derivative terms have been omitted. From the above convective wave equation, we can deduce following the same procedure two transport equations, one forward and one backward.

\textit{Wave equation with nested convolution kernels.} To give a different example, we may consider the following kinetic equation proposed by Zwanzig and Nakajima to describe open non-equilibrium systems with memory effects \cite{Zwanzig},

\begin{equation}
\left(\frac{d }{dt}-\hat{\mathcal{P}}\hat{L}\right)\rho_{rel}(t)=\int _{0}^{t} dt'\hat{\mathcal {K}}(t')\rho_{rel}(t-t'),
\end{equation}

\noindent
where $\hat{\mathcal{P}}$ the projection operator, $\hat{L}$ the Liouville superoperator, $\hat{\mathcal{K}}$ the memory superoperator kernel and $\rho _{rel}$ the relevant part of the density operator. On this occasion we multiplying the above with $\frac{d }{d t}+\hat{\mathcal{P}}\hat{L}$ gives

\begin{equation}
\left(\frac{d ^{2}}{d t^{2}}-\hat{\mathcal{P}}\hat{L}^{2}\right)\rho_{rel}(t)=  \int _{0}^{t} dt'   \hat{ \mathcal {J} }(t')\rho_{rel}(t-t').
\end{equation}

\noindent
where

\begin{equation}
   \hat{ \mathcal {J} (t')}=\hat{\mathcal{P}}\hat{L}\hat{\mathcal {K}}(t')+2\hat{\mathcal {K}}(t')\hat{\mathcal{P}}\hat{L} +\hat{\mathcal {K}}(t')\int _{0}^{t'} dt''\hat{\mathcal {K}}(t''),
\end{equation}

\noindent
We notice the quadratic Liouvillian term on the left-hand side, which could lead to non-classical correlations and spatial non-locality and the nested memory kernel on the right-hand one that may lead to finite duration measurements. Taking the Laplace transform $\mathcal {L}$ of $\hat{\rho}_{rel}$,

\begin{equation}
   \tilde{ \rho}_{rel}(s)=\int _{0}^{\infty}dt e^{-st}\rho_{rel}(t) ,
\end{equation}

\noindent
we obtain from (46) after multiplying with $e^{+st}$ and integrating

\begin{equation}
    \rho_{rel}(t)= \frac{1}{2\pi i}\int _{\gamma-i\infty}^{\gamma+i\infty}ds e^{st}\frac{[s\rho_{rel}(s) - \rho 
_{rel}(0)] }{s^{2}-\hat{\mathcal{P}}\hat{L}^{2}+\mathcal{L}\{\hat{ \mathcal {J}(t')} \}},
\end{equation}

\noindent
 which at least in principle could be evaluated and so extract the relevant part of the density operator. 
 
Using the same line of reasoning one may extract an analogous equation from classical or quantum kinetic reaction-diffusion ones such as the Fokker-Planck or Burgers' equation. It is also possible to establish a cluster of hierarchy equations that relate the distribution functions or Green functions of different numbers of particles \cite{Leeuwen}.

\section{Damping and non-local dispersion effects}

 \textit{Convective wave equation for the spatial probability density} We now repeat the same procedure to derive an equation for the spatial probability density. Starting from the continuity equation

\begin{equation}
\frac{\partial \rho}{\partial t}+\sum _{i=1}^{N} \mathbf{v}_{i}\nabla _{i} \rho=-\rho \sum _{i=1} ^{N} \nabla _{i} \mathbf {v}_{i},
\end{equation}

\noindent
and using the total time derivative is also written as

\begin{equation}
\dot{\rho}=-\rho \sum _{i=1}^{N}\nabla_{i}\cdot \mathbf{v}_{i},
\end{equation}

\noindent
with a solution of the form

\begin{equation}
    \rho (\mathbf{r},t)=  e^{-\int ^{t}_{t_{0}} d\tau \nabla\cdot \mathbf{v}} \rho  (\mathbf{r}_{0},t_{0}).
\end{equation}

\noindent
Applying on (50) the operator $\frac{\partial}{\partial t}-\sum _{i=1}^{N} \mathbf{v}_{i}\cdot\nabla _{i}$
we obtain

\begin{equation}
    \ddot{\rho}-2\mathbf{v}\cdot\nabla \dot{\rho}=-\dot{\rho}\nabla\cdot \mathbf{v}-\rho \nabla\cdot \dot{\mathbf{v}}+2\mathbf{v}\cdot \nabla \left(\rho\nabla\cdot\mathbf{v}\right),
\end{equation}

\noindent
and finally

\begin{equation}
   \dot{\rho}=\pm \sqrt{2\int d\rho \left[\frac{1}{m} \rho\sum _{i=1}^{N}\nabla_{i}^{2}(V+Q)-\dot{\rho}\sum _{i=1}^{N}\nabla_{i}\cdot \mathbf{v}_{i}+C\right]}.
\end{equation}

\noindent
Approximately this is written as

\begin{equation}
    \ddot{\rho} \approx \rho \left [\left(\nabla\cdot \mathbf{v}\right)^{2}-\nabla\cdot \dot{\mathbf{v}}\right],
\end{equation}

\noindent
with a solution

\begin{equation}
   \pm \int \frac{d\rho}{\sqrt{C_{1}+\int d\rho \rho \left [\left(\nabla\cdot \mathbf{v}\right)^{2}-\nabla\cdot \dot{\mathbf{v}}\right]}}\approx t+C'.
   \end{equation}

\noindent
This type of time evolution may be responsible for intrinsic loss of coherence and correspond to non-exponential decay or amplification processes and in a certain region exponential ones with a spatio-temporal profile rate

\begin{equation}
 \Gamma(\mathbf{r},t)\sim\mp\sqrt{\frac{1}{m}\left [\frac{\partial}{\partial t}\nabla S+\frac{1}{m}\nabla S\cdot \nabla(\nabla S)\right]} ,  
\end{equation}

\noindent
which means that for large masses the first term is the dominant one whereas for small ones it is the first one.

Since (54) is first-order in time we can use it to find expressions for the rate of the total probability $\int d^{3}x \rho$. The probability density can be expressed concerning the Wigner function as the marginal integral of the momentum, for the single-particle case in three dimensions is written as $\rho (\mathbf{r},t)=\int d\mathbf{p}  W (\mathbf{r},\mathbf{p},t)$, by which we arrive at

\begin{equation}
    \frac{dP}{dt}= \mp \int d^{3}x 
\int d\mathbf{p}'\sqrt{\mathcal{G}},
\end{equation}

\noindent
where 

\begin{equation}
  G= \int dW \Bigg\{\frac{4\mathbf{p}}{m}\nabla\dot{W}+2\int d^{N}\mathbf{p}'\Bigg[\left( \dot{\tilde{V}}-\frac{2\mathbf{p}}{m}\cdot\nabla\tilde{V}\right)
-\tilde{V}\left(\frac{\mathbf{p}}{m}\cdot\nabla+\int d^{N}\mathbf{p}' \tilde{V} \right)\Bigg]W\Bigg\}
\end{equation}

\noindent
In the same manner and rate of entropy production is given by

\begin{equation}
    \frac{dS}{dt}= k\int d^{3}x \int d\mathbf{p}' \left(1+ln\int d\mathbf{p}'W \right)\mathcal{G}.
\end{equation}

\noindent
To quantify the measure of disorder we define the Fisher information 

\begin{equation}
    \mathcal {I}=\int d^{3}x \frac{(\nabla \rho)^{2}}{\rho},
\end{equation}

\noindent
and obtain its rate of change as

\begin{equation}
    \frac{d\mathcal{I}}{dt}=-\int d^{3}x \int d\mathbf{p}'\frac{\left[\left(\nabla \int d\mathbf{p}' W (\mathbf{r},\mathbf{p},t)\right)^{2}-2\nabla \int d\mathbf{p}' W (\mathbf{r},\mathbf{p},t)\nabla \right]\mathcal{G}}{(\int d\mathbf{p}' W (\mathbf{r},\mathbf{p},t))^{2}}.
\end{equation}

\noindent
These non-unitary phenomena may be associated with decay or amplification and they are not extrinsic or contingent theoretical elements but rather nomological propositions. From another point of view, they belong to a conceptual framework put forward by de Broglie relevant to the thermodynamics of an isolated particle and imply a hidden sub-quantum reservoir \cite{deBroglie}. Their origin lies in the differentiation we have performed, in other words, we could have added or subtracted a source-sink term in (13) as

\begin{equation}
   \dot{W}= \left[\frac{\partial }{\partial t}+\left(\sum_{i=1}^{N}\frac{\mathbf{p}_{i}}{m_{i}}\cdot\nabla_{i}\right)\right]W=\int d^{N}\mathbf{p}'\tilde{V} W+f(\mathbf{r}_{1},\mathbf{r}_{2},..,\mathbf{r}_{N},t)
\end{equation}

\noindent
and yet also arrive at the same expression (16) and consequently (21) provided that
satisfies the constraint

 \begin{equation}
\left(\frac{\partial}{\partial t}-\sum_{i=1}^{N}\frac{\mathbf{p}_{i}}{m_{i}}\nabla_{i}\right)f(\mathbf{r}_{1},\mathbf{r}_{2},..,\mathbf{r}_{N},t)=-\int d^{N}\mathbf{p}'\tilde{V} W, 
\end{equation}   

\noindent
and we may identify $f(\mathbf{r}_{1},\mathbf{r}_{2},..,\mathbf{r}_{N},t)$ by setting (61) against the square root of (54). It follows from (63) then that probability is not conserved locally. 

We now adopt accordingly a procedure familiar in plasma physics and specifically Landau damping \cite{Landau}, namely taking initially a Fourier transform in space and consecutively a Laplace transform in time for the Wigner function in one dimension for a single particle 

\begin{equation}
    {W}(k,t)=\frac{1}{2\pi}\int _{-\infty}^{+\infty}W(x,t)e^{-ikx}dx,
\end{equation}

\begin{equation}
{f}(k,t)=\frac{1}{2\pi}\int _{-\infty}^{+\infty}f(x,t)e^{-ikx}dx.
\end{equation}

\noindent
where

\begin{equation}
f(x,t) \approx  \pm \sqrt{2\int dW  \int d{p}' \dot{\tilde{V}}W }  \left(1+\frac{\int dW\int dp'\tilde{V}\int d{p}' \tilde{V}W}{ 2\int dW\int dp'\dot{\tilde{V}}W}\right) -\int d{p}' \tilde{V}W .
\end{equation}

\noindent
Taking the Fourier transform of (63) leads to

\begin{equation}
    \frac{\partial W(k,t)}{\partial t}+ik\frac{p}{m}W(k,t)=f(k,t)+\int \tilde {V}(k,t)*W(k,t)dp ,
\end{equation}

\noindent
where the last term is a convolution one. Then writing the expressions for the Laplace transforms of

\begin{equation}
   \tilde W(s,t)=\int _{0}^{+\infty}{W}(k,t)e^{-st}dt,
\end{equation}

\begin{equation}
  \tilde  f(s,t)=\int _{0}^{+\infty}{f}(k,t)e^{-st}dt,
\end{equation}

\noindent
multiplying by $e^{-st}$ and after integrating we get

\begin{equation}
   \left(s+ik\frac{p}{m}\right)\tilde W(s,t)=\tilde W(s,0)+ \tilde f(s,t)+\int \tilde {V}(s,t)*W(s,t)dp
\end{equation}

\noindent
from which it follows

\begin{equation}
 \tilde W(s,t)=\frac{\left(s-ik\frac{p}{m}\right)\left[\tilde{W}(s,0)+\int \tilde {V}(s,t)*W(s,t)dp\pm\frac{1}{\pi}\int _{0}^{+\infty}dte^{-st}\int _{-\infty}^{+\infty}dxe^{-ikx}f(x,t)\right]}{\left(s^{2}+k^{2}\frac{p^{2}}{m^{2}}\right)},
\end{equation}

\noindent
and

\begin{equation}
    W(x,t)=\frac{1}{4\pi ^{2}i}\int _{-\infty}^{+\infty}dxe^{ikx}\int _{\gamma-i\infty}^{\gamma+i\infty}\tilde{W}(s,t)e^{st}ds.
\end{equation}

\noindent
This Lorentzian form of the integral dispersion equation along with the imaginary component signifies damping. Inverting the Fourier and Laplace transforms would solve the problem in principle even though in practice the calculations involved may be quite tedious. 

Since the above is a rather formal expression, a more convenient way perhaps to deal with the problem of dispersion when we include interactions is to make use of (16) and (17) and resort to approximate values of the Fourier integral. Expressing $W(x_{1},x_{2},p_{1},p_{2},t)$ as a wavepacket of different frequencies in one dimension

\begin{equation}
W_{12}(x_{1},x_{2},t) = \frac{1}{\sqrt{2\pi}}\int _{-\infty}^{+\infty}A(k_{1},k_{2})e^{i[k_{1}x_{1}+ik_{2}x_{2}-\omega(k_{1},k_{2})t]}dk_{1}dk_{2}.
\end{equation}

\noindent
Near the dominant frequency $k_{0}=(k_{1}(0),k_{2}(0))$ according to the stationary phase approximation the integral at large times is written as \cite{Billingham, Bruijn}

\begin{equation}
  W_{12} \approx A(k_{0})\frac{2\pi e^{i[k_{0}x-\omega(k_{0})t\mp \frac{\pi}{2}]}}{t\sqrt{\frac{\partial ^ {2}\omega (k_{0}) }{\partial k_{1}^{2}}\frac{\partial ^ {2}\omega (k_{0}) }{\partial k_{2}^{2}} -\frac{\partial ^ {2}\omega (k_{0)} }{\partial k_{1}k_{2}}} },
\end{equation}

\noindent
writing the frequency as $\omega=\omega_{R}+i\omega_{i}$ we get

\begin{equation}
    \omega_{R}^{2}= \left(\frac{p_{1}k_{1}}{m_{1}}+\frac{p_{2}k_{2}}{m_{2}}\right)^{2}+h,
\end{equation}

\noindent
where

\begin{equation}
    h=\int dp_{1}' dp_{2}'\left[\tilde{V}\int  dp_{1}' dp_{2}'\tilde{V}- \left(\frac{\partial \tilde{V}}{\partial t}-\frac{p_{1}}{m_{1}}\frac{\partial \tilde {V}}{\partial x_{1}}-\frac{p_{2}}{m_{2}}\frac{\partial \tilde {V}}{\partial x_{2}}\right)\right].
\end{equation}

\noindent
The imaginary part of the frequency is given by

\begin{equation}
    \omega _{I}=\frac{\left(\frac{p_{1}k_{1}}{m_{1}}+\frac{p_{2}k_{2}}{m_{2}}\right)\int dp_{1}'dp_{2}' \tilde{V}}{\omega _{R}},
\end{equation}

\noindent
which can be negative, so the frequency $\omega$ may have a pole at the lower half-plane which in turn means that the conditions for the Kramers-Kronig relations may be violated which means that the link between causality and the analyticity of complex frequency does not hold in general. This complex frequency also naturally points to damping effects. The group velocity for the real frequency part is

\begin{equation}
    \frac{d\omega _{R}}{dk_{1,2}}= \frac{\pm \left(\frac{p_{1}k_{1}}{m_{1}}+\frac{p_{2}k_{2}}{m_{1}}\right)\frac{p_{1,2}}{m_{1,2}}}{\sqrt{
    \left(\frac{p_{1}k_{1}}{m_{1}}+\frac{p_{2}k_{2}}{m_{2}}\right)^{2}+h}},
\end{equation}

\noindent
 which is non-local apart from non-separable; interestingly, the imaginary group velocity is local. The physical implications of this section are related to the group velocity of information propagation and suggest superluminal but not necessarily instantaneous signals.

\section{Concluding remarks}
The present paper is mainly concerned with the evolution of wave equations in phase and configuration space, with the distinctive feature of being non-separable, and the resulting propagation of correlations. Proceeding from the transport Wigner equation we have produced firstly a non-separable equation second-order in time, that involves inter-site spatial derivatives. Initially having employed the free Wigner we derived a partial derivative equation and after including interactions an integro-differential one. Consequently, and with the help of the many-particle total derivative we formally reduced it again to a set of two distinct kinetic equations and gave expressions for the total probability rate of probability gain or loss and the entropy production. These non-unitary processes are intrinsic to formalism and not added in a phenomenological manner or result from interaction with a thermal bath and the damping constant depends on the hydrodynamic particle velocity. Finally, the occurrence of dispersion in a bi-particle setting was investigated by using a variant of the wave equation of motion which illustrated the possibility of negative propagated probability densities, non-local group velocities and quadrupole effects, along with a wave equation with nested convolution memory effects for the density operator.

\end{document}